\documentclass[conference, a4paper]{IEEEtran}
\IEEEoverridecommandlockouts
\usepackage{cite}
\usepackage{amsmath,amssymb,amsfonts}
\usepackage{algorithmic}
\usepackage{graphicx}
\usepackage{textcomp}
\usepackage{xcolor}
\usepackage{multirow}
\def\BibTeX{{\rm B\kern-.05em{\sc i\kern-.025em b}\kern-.08em
    T\kern-.1667em\lower.7ex\hbox{E}\kern-.125emX}}
\begin{document}

\title{Enhanced Security against Adversarial Examples Using a Random Ensemble of Encrypted Vision Transformer Models\\
}

\author{\IEEEauthorblockN{1\textsuperscript{st} Ryota Iijima}
\IEEEauthorblockA{\textit{Tokyo Metropolitan University} \\
Tokyo, Japan \\
iijima-ryota@ed.tmu.ac.jp}
\and
\IEEEauthorblockN{2\textsuperscript{nd} Miki Tanaka}
\IEEEauthorblockA{\textit{Tokyo Metropolitan University} \\
Tokyo, Japan \\
miikeneko1221@outlook.com }
\and
\IEEEauthorblockN{3\textsuperscript{rd} Sayaka Shiota}
\IEEEauthorblockA{\textit{Tokyo Metropolitan University} \\
Tokyo, Japan \\
sayaka@tmu.ac.jp}
\and
\IEEEauthorblockN{4\textsuperscript{th} Hitoshi Kiya}
\IEEEauthorblockA{\textit{Tokyo Metropolitan University} \\
Tokyo, Japan \\
kiya@tmu.ac.jp}
% \and
% \and
}

\maketitle

\begin{abstract}
Deep neural networks (DNNs) are well known to be vulnerable to adversarial examples (AEs).
In addition, AEs have adversarial transferability, which means
AEs generated for a source model can fool another black-box model (target model) with a non-trivial probability.
In previous studies, it was confirmed that the vision transformer (ViT) is more robust against the property of adversarial transferability than convolutional neural network (CNN) models such as ConvMixer, and moreover encrypted ViT is more robust than ViT without any encryption.
In this article, we propose a random ensemble of encrypted ViT models to achieve much more robust models.
In experiments, the proposed scheme is verified to be more robust against not only black-box attacks but also white-box ones than convention methods.
\end{abstract}


\section{Introduction}
\noindent
Deep neural networks (DNNs) have been developed in various fields, but they have critical problems to be resolved \cite{hitoshi2022an}.
One of the problems is that DNNs are vulnerable to adversarial examples (AEs), so a trained model is fooled by using AEs.
In addition, AEs also have a property, called the transferability of AEs, which means that AEs designed for a model (source model) fool a black-box model (target model) with a non-trivial probability as well as the source model.
In this paper, we aim to construct robust models against AEs including the transferability of AEs. \par
To achieve robust models against AEs, various studies have been reported so far \cite{pang2019improving,yang2020dverge,aprilpyone2021block,april2022priavcy,maung2020encryption,aprilpyone2021protection,maungmaung2021ensemble}.
In previous studies, it was confirmed that the use of models trained with encrypted images is robust against white-box attacks, but it is not effective under state-of-the-art black-box attacks \cite{aprilpyone2021block,april2022priavcy,maung2020encryption,aprilpyone2021protection,maungmaung2021ensemble}.
The vision transformer (ViT) was also demonstrated to be more robust against the property of adversarial transferability than convolutional neural network (CNN) models such as ConvMixer, and moreover encrypted ViT is more robust than ViT without any encryption \cite{tanaka2022on}. \par
Because of such a situation, in this paper, we propose a random ensemble of encrypted ViT models to achieve much more robust models.
In experiments, the proposed scheme is verified to be more robust against not only black-box attacks but also white-box ones than convention methods.

\section{Related Work}
\subsection{Vision Transformer}
The vision transformer (ViT) \cite{dosovitskiy2021an} is known as a model that provides high performance in classification tasks.
Figure 1 shows the architecture of ViT.
ViT classifies images according to the following steps.
\begin{enumerate}
    \item Split an image into fixed-size patches, and linearly embed each of them.
    \item Add position embedding to patch embedding.
    \item Feed the resulting sequence of vectors to a standard transformer encoder.
    \item Feed the output of the transformer to a multi-layer perceptron (MLP), and get a result.
\end{enumerate}
ViT is usually used after fine-tuning a pre-trained model.
In previous studies, it was shown that fine-tuning by using encrypted images improves the robustness against AEs \cite{tanaka2022on}.
In this paper, we also use ViT models fine-tuned with encrypted images as sub-models for a random ensemble.

\begin{figure}
    \centering
    \includegraphics[scale=0.4]{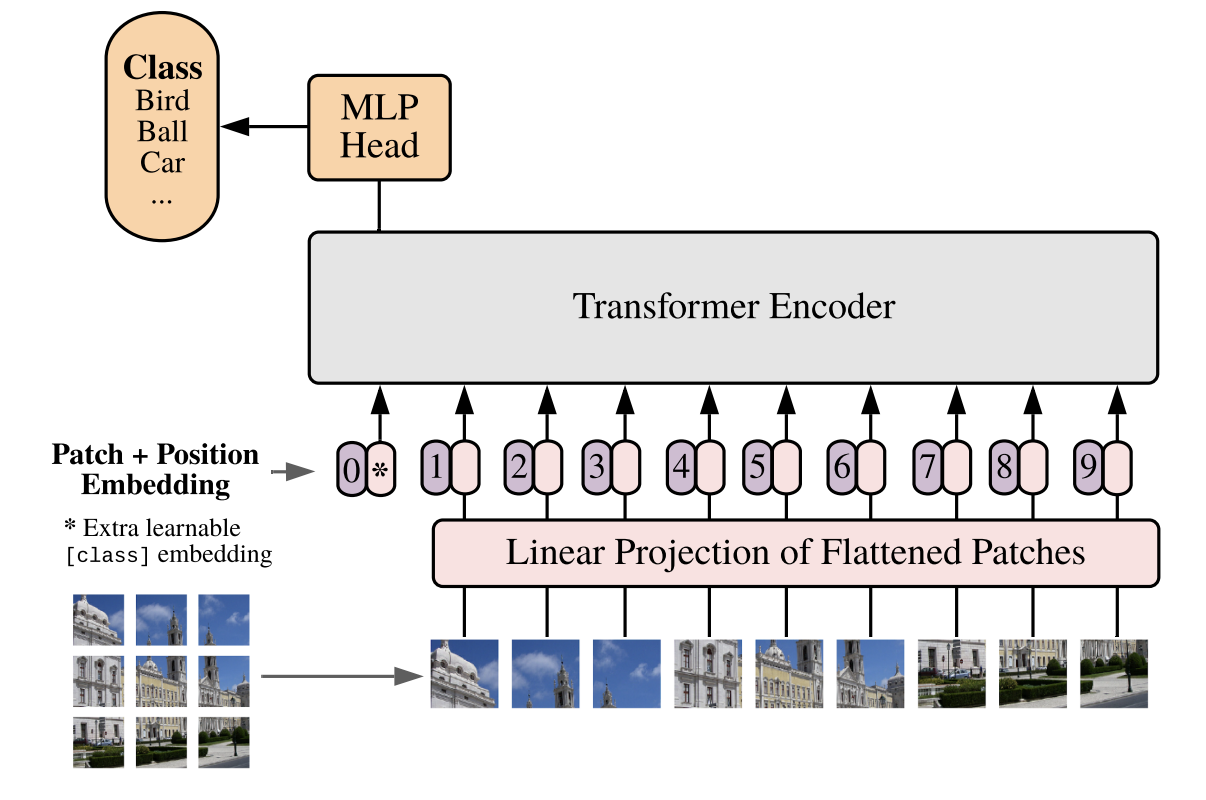}
    \caption{Architecture of Vision Transformer \cite{dosovitskiy2021an}}
    \label{fig:fig1}
\end{figure}

\subsection{Adversarial Examples}
Depending on the ability of adversaries, there are two types of attacks: white-box attacks \cite{goodfellow2015explaining,madry2018towards,carlini2017towards,moosavi2016deepfool} and black-box attacks \cite{andriushchenko2020square}.
The adversaries have complete knowledge of the target model and data information in white-box settings.
In contrast, in black-box settings, adversaries can transfer the generated AE to the unknown deployed model based on AE transferability.
Furthermore, AEs can be categorized into two types in terms of the goal of adversaries.
Target attacks mislead the output of models to a  specific class.
In contrast, non-targeted attacks aim to mislead models to an incorrect class. \par
AutoAttack \cite{croce2020reliable} was proposed to evaluate the robustness of defense methods against AEs in an equitable manner.
The attack method consists of four parameter-free attack methods: Auto-PGD-cross entropy (APGD-ce), Auto-PGD-target (APGD-t), FAB-target (FAB-t) \cite{croce2020minimally}, and Square attack \cite{andriushchenko2020square}.
In this paper, we use APGD-ce and Square attack as a white-box attack and a black-box attack to evaluate an random ensemble, respectively.
Both attacks are non-targeted ones. \par
Adversarial training \cite{goodfellow2015explaining,kurakin2017adversarial, Hongyang2019theoratically,carmon2019unlabeled} is widely known as a defense method against AEs, where AEs are used as training data to improve the robustness against AEs.
However, it degrades the performance of models when clean
images are input.
Defense methods against AEs are expected to meet the following requirements in general.
\begin{itemize}
    \item No performance degradation even when clean images are input.
    \item Being robust enough against all attack methods.
\end{itemize}

\section{Proposed Method}
\subsection{Overview}
Figure 2 shows the framework of the proposed scheme.
At first, a provider trains $N$ sub-models with images encrypted with secret keys $K=\{K_{1}, \dots, K_{N} \}$.
Next, the provider constructs a random ensemble of the sub-models as an image classifier.
The provider encrypts a test image with $K$ to generate $N$ encrypted test images, and the encrypted images is input to the classifier (a random ensemble of encrypted ViT models) to get an estimate result.

\begin{figure}
    \centering
    \includegraphics[scale=0.6]{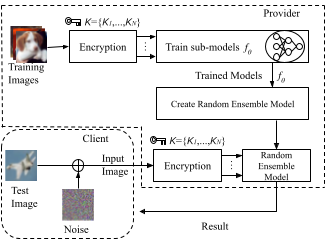}
    \caption{Framework of proposed scheme.}
    \label{fig:fig2}
\end{figure}

\subsection{Random ensemble of sub-models}
Figure 3 shows the details of a random ensemble of $N$ encrypted sub-models.
Every sub-model is ViT, and a different secret key is assigned to each sub-model for image encryption.
In this paper, pixel shuffling is used for image encryption as in \cite{tanaka2022on}.
The following steps are carried out to generating encrypted images for pixel shuffling.
\begin{enumerate}
    \item Split an image into non-overlapped blocks with a size of $M \times M$, where $M$ is the same size as the patch size of ViT.
    \item Flatten each block into a vector.
    \item Randomly permute pixels in each vector to generate an encrypted vector by using key $K_i, i=1,2,dots,N$.
    \item Rebuild the encrypted vector into the encrypted block.
    \item Concatenate the encrypted blocks into an encrypted image.
\end{enumerate}
Please note that $N$ encrypted images are generated from an image by using $N$ keys, and any clients do not know the keys.
In the proposed method, $S$ outputs are randomly selected from $N$ outputs of sub-models where $3 \leq S \leq N$.
The final outputs are determined by the average of $S$ outputs.

\begin{figure}
    \centering
    \includegraphics[scale=0.24]{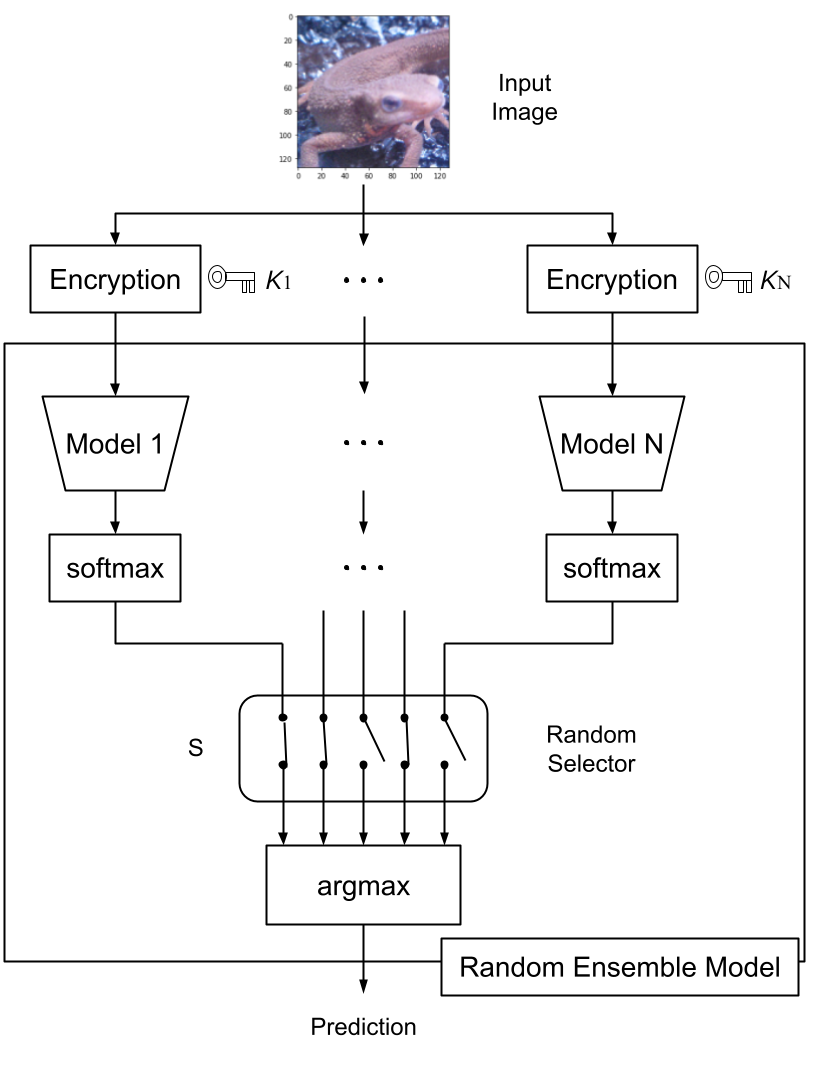}
    \caption{Random ensemble of encrypted models.}
    \label{fig:fig3}
\end{figure}

\section{Experimemt}
\subsection{Experimental Setup}
Experiments were conducted on the CIFAR-10 dataset.
The dataset, which consists of $60,000$ images with size $32 \times 32 \times 3$, was divided into $50,000$ and $10,000$ images for fine tuning and testing, respectively.
All images were resized to $224 \times 224 \times 3$ to fit the input to ViT and scaled to $[0,1]$ as a range of the values.
We used finetuned ViT models with a patch size of $P=16$ where ViT was pre-trained with ImageNet-21k \cite{dosovitskiy2021an}.
ImageNet-21k is a dataset consisting of $21,000$ classes with a total of $1,400$ million patches, which were resized to image size $224 \times 224 \times 3$ when pre-training ViT. For fine tuning, a learning rate of $lr=0.03$ was set, and we ran $5,000$ epochs.
For model encryption, a block size of $M=16$ was used as well as the patch size of ViT .
The robustness models were evalutaed by using two attack methods, APGD-ce, which is included in AutoAttack \cite{croce2020reliable} as a white-box attack, and Square attack \cite{andriushchenko2020square}, which is also included in AutoAttack as a black-box attack.
The Attack Success Rate (ASR) was used as an evaluation metric.

\subsection{Experimental Result}
First, we compared the proposed random ensemble models with ensemble models (no random selection) under the use of APGD-ce and Square attack (see Table \ref{tab:tab1}), where both models consisted of $N = 4$ sub-models encrypted by using different keys.
From Table \ref{tab:tab1}, the proposed ensemble outperformed the conventional one under Square, but both ensembles had almost the same ASR values under APGD-ce where it was assumed that an attacker knew all keys of sub-models. \par
In Table 2, Next, the ASR values of random ensemble models against APGD-ce were evaluated when the number of keys known to an attacker was varied.
As show in Table \ref{tab:tab2}, APGD-ce attacks failed when an attacker did not know any keys or knows only one key.

% after
\begin{table}[t]
    \centering
    \caption{Comparison of Ensemble Model and Random Ensemble Model with ASR (all keys are known to an attacker)}
    \begin{tabular}{c|cc}
    \hline
    \multirow{2}{*}{Source Model} & \multicolumn{2}{c}{Attack Method} \\
    \cline{2-3}
    &  APGD-ce & Square \\
    \hline
    Ensemble & 100.00 & 98.43 \\
    \hline
    Random Ensemble &  \multirow{2}{*}{99.90} & \multirow{2}{*}{\textbf{22.21}} \\
    (Proposed) & & \\
    \hline
    \end{tabular}
    \label{tab:tab1}
\end{table}

\begin{table}[t]
    \centering
    \caption{ASR of Random Ensemble Model (some keys are know to an attacker)}
    \begin{tabular}{c|c|c}
    \hline
    \multicolumn{2}{c|}{Source Model} & Attack Method \\
    \hline
    Model & \# leaked keys & APGD-ce \\
    \hline
    \multirow{4}{*}{Ensemble}& 4 & 100.00 \\
    & 2 & 97.25 \\
    & 1 & 2.24 \\
    & 0 & 0.04 \\
    \hline
    & 4 & 99.90 \\
    Random Ensemble & 2 & 71.74  \\
    (Proposed) & 1 & 2.73  \\
    & 0 & 0.12 \\
    \hline
    \end{tabular}
    \label{tab:tab2}
\end{table}

\section{Conclusion}
The use of encrypted models was known to be effective against white-box attacks if secret keys are not open.
In this paper, we proposed an novel method with encrypted models, which is carried out on the basis of an random ensemble of encrypted ViT models, so that the robustness of models is enhanced against black-box attacks in addition to against white-box attacks
when disclosing a few keys.

\section*{Acknowledgments}
This study was partially supported by JSPS KAKENHI (Grant Number JP21H01327) and JST CREST (Grant Number JPMJCR20D3).

\bibliographystyle{IEEEtran}
\bibliography{IEEEabrv,main}

% \vspace{12pt}
% \color{red}
% IEEE conference templates contain guidance text for composing and formatting conference papers. Please ensure that all template text is removed from your conference paper prior to submission to the conference. Failure to remove the template text from your paper may result in your paper not being published.

\end{document}